# Bosonic mode interpretation of novel STM and related experimental results, within boson-fermion modelling of HTSC

**John A Wilson**

H.H. Wills Physics Laboratory,
University of Bristol,
Tyndall Avenue,
Bristol BS8 1TL.   U.K.

**Abstract**

This paper seeks to synthesize much recent work on the HTSC materials around the latest energy resolved scanning tunnelling microscopy (STM) results from Davis and coworkers. The conductance diffuse scattering results in particular are employed as point of entry to discuss bosonic modes, both of condensed and uncondensed form. The bosonic mode picture is essential to understanding an ever growing range of observations within the HTSC field. The work is expounded within the context of the negative-*U*, boson-fermion modelling long advocated by the author. This general approach is presently seeing much theoretical development, into which I have looked to couple many of the experimental advances. While this formal theory is not yet sufficiently detailed to cover adequately all the experimental complexities presented by the real cuprate systems, it is clear it affords very appreciable support to the line taken. An attempt is made throughout to clarify why and how it is that these novel phenomena are tied so very closely to this particular set of materials.

e-mail: john.a.wilson@bristol.ac.uk

**§1. Introduction to the energy resolved STM work of Hoffman *et al*.**

Recently Hoffman *et al* [1] have uncovered in an energy resolved 4K STM study of local tunnelling conductance into Bi-2212 that over the energy range 5 to 30 meV additional novel scattering information is forthcoming. Through a Fourier transformation of their real space local conductance maps they show that structured incommensurate elastic scattering of the incoming quasiparticles is being incurred. This scattering is quite distinct from that previously reported due to (*i*) low energy impurity effects, (*ii*) magnetic effects, (*iii*) charge stripe effects, (*iv*) vortex effects, and (*v*) the Bi-O layer superlattice modulation. The newly found modulations in the tunnelling conductance signal are much more diffuse than those from (*v*), and the action at the wavevectors involved is seen besides to be strongly energy dependent. The dispersion characteristics manifest in the new scattering are moreover not those of the antiferromagnetic magnons of low doping, nor likewise of the lattice phonons. The resolved wavevectors ($\mathbf{q}_A$ and $\mathbf{q}_B$) are found to relate directly to the basal Fermi surface geometry, $\mathbf{k}_F$, and the scattering is coupled furthermore with tip binding energies roughly equal in value to the superconducting gap. As a result Hoffman *et al* have tentatively presented their 4K STM conductance data in terms of the introduced electrons being scattered from the Bogoliubov quasiparticles. The density of states peak for the latter then should of course map out the energy below $\mathbf{k}_F$ of the bottom of the superconducting band gap as structured by the $d_{x2\text{-}y2}$ symmetry HTSC order parameter $\Delta(\mathbf{k}_F)$. However upon closer examination of the STM results we will show that while at each level of sample hole doping reported the locus from the full set of conductance modulation wavevectors indeed creates a shape in *k*-space very close to the ARPES-determined normal state Fermi surface, the associated binding energies $E_q(\theta)$ are consistently significantly greater than the gap $\Delta(\theta)$ but less than $2\Delta(\theta)$ – prior at least to the 'hot spot' being reached (see below). It will be argued that this observation is in accord with the existence of a dispersed uncondensed boson mode, as figures in the negative-*U* two-subsystem treatment of HTSC from the current author [2]. Formal development in modelling of the mixed boson-fermion and/or negative-*U* form recently has been much expanded by Micnas *et al* [3], by Domanski *et al* [4], and by Casas, de Llano, Solis *et al* [5], closer examination being made of questions relating in particular to the electronic specific heat, entropy and condensation energy data from Loram *et al* [6] and to the Uemura type rendering of the μSR data [7].

A matter that we will return to in due course but one which at this point must be registered is the misleading reading of the true value of the superconducting gap from the ARPES spectra adopted by Hoffman *et al*. In the latter spectra the leading peak has to be viewed - unlike in earlier type readings of such data, and taken up now in [1] - as much more closely defining $2\Delta$ than it does $\Delta$. $\Delta$ in fact is better monitored by the inflexion point of the leading edge. At optimal doping in YBCO-123 and BSCCO-2212 it is the maximal gap values $2\Delta_o$ which fall just short of 40 meV. Thermally activated experimentation such as specific heat and nmr, along with phonon line-width and electronic Raman analyses [2e (§E16) and 2f (ref 36)] all concur on such a $2\Delta_o$ value ($\cong$ 320 $cm^{-1}$). This yields then the well-known strong-coupling $2\Delta_o/kT_c$ ratio of approximately 5.5. A proper reading of $2\Delta$ directs interpretation of what Hoffman *et al*'s new results mean along significantly different lines from those followed in [1]. The proper interpretation of the ARPES spectrum from Bi-2212 has been a long and complex task – one involving the joint effects of the local boson mode and bilayer splitting, in addition to matters of quasiparticle delocalization.

What is most striking about the new STM results, as indicated above, is that for any given doping level the *maximal* (*) scattering signal identified in the conductance occurs at some energy $E_q$* appreciably less than 2Δ(**k***,*p*',*T* = 0), for coupled $\mathbf{q}_A$*/$\mathbf{q}_B$*. The latter wavevectors specify a set of equivalent points within *k*-space which fall well removed from the (π,0) saddles in the band structure, and likewise from the renowned 'hot spots' sited on lines (0,π) to (π,0), *etc*. where those lines intersect the Fermi surface. In the spin fluctuation interpretation of HTSC the 'hot spots' are to be associated with the highly characteristic (π,π) scattering physics [8], preeminently in evidence in the striking 'resonance mode' of inelastic neutron scattering, where the scattering is normally treated in standard nesting fashion [9]. By contrast our own negative-*U* interpretation of the π,π scattering involves not true *magnetic* spin excitation but a singlet-triplet spin flip pair breaking of local pairs [2]. The maximal scattering in the new STM data, besides arising in k-space as close to the $d_{x2-y2}$ nodes as to the hot points, does so at a binding energy $E_q$* which as noted above is only ≈ ½.2$\Delta_o$(**k***,*p*'). It would appear that the STM conductance signal becomes maximized where/when the dispersed boson mode of the postulated negative-*U* model stands not too severely perturbed either by pair formation or pair dissolution, these in the model coming in the vicinity respectively of the hot spots and of the gap nodes.

The above STM scattering results are employed in the present paper as the point of departure against which an extensive range of recent experimental results can be addressed within the framework of the author's long-standing negative-*U* interpretation of HTSC behaviour. Section 2 presents some of the earlier detailing of the model along with a full description of the new STM scattering experiment and data. Comparison is made with recent ARPES results. Section 3 presents the STM scattering entity as being a dispersed mode of uncondensed local pair bosons, and it contrasts the STM results with what has been recorded in inelastic neutron scattering events, both of spin-flip and phononic issue. Section 4 deals conversely with the condensed boson state and with its excited plasma mode. The latter is presented as responsible for the kinking induced in the quasiparticle bands around 60 meV, and also for the strong HREELS signal found at this energy. The very recent Nernst results are considered here and again at the beginning of Section 5. Section 5 in the main deals with the nature and consequences of the mixed-valent inhomogeneity, returning to further new STM work. Section 6 starts again from STM work in introducing the variety of roles LO phonons play in HTSC phenomena, and in particular $\Sigma_1$ coupling to the plasma mode of the boson condensate. The consequences for phonon dispersion, Raman, IR, nmr, isotopic shift and high pressure results are all detailed. Finally Section 7 considers more closely the current theoretical state of play regarding the nature of HTSC in the cuprates. It incorporates the effect of a slight breakage of e-h symmetry in the Bogoliubov sense, and it contrasts this with the strong asymmetry prevalent between hole and electron carrier types within the bank of known superconductors. Shell-filling effects are contrasted with those of the band Jahn-Teller type as regards sourcing of exotic non-BCS type superconductivity. Throughout indication is given of how spin-fluctuation, phonon and Marginal Fermi Liquid modellings of HTSC each fall short of what can be accomplished by pursuing the boson-fermion, resonant negative-*U*, two-subsystem approach.

**§2. Relationship of the new STM data with 2Δ(θ) gap data from ARPES.**

Hoffman *et al* observe that the new diffuse peaks appearing in the (**r**:**k**) Fourier analysis of the STM conductance data clearly seems to reflect the geometry of the Fermi surface. In fact the two wavevectors **q** detailing the scattering events in *k*-space seem in elastic fashion to span at selected energy *E* below the Fermi surface between equivalent degenerate points within the zone, set $\mathbf{q}_A$ lying parallel to the Cu-O bonds and set $\mathbf{q}_B$ directed at 45$^o$ to these. The $\mathbf{q}_A$ span across the occupied arms of the Fermi sea while the $\mathbf{q}_B$ span between these arms across the unoccupied parts of the zone. Accordingly note that the $\mathbf{q}_B$ lie parallel to the Bi-O superlattice modulation, sharp diffraction spots from which are very visible in the STM conductance F.T.'s [1; fig 3]. The above assignment is in full accord with the observation made in [1] that as the (hole) doping is augmented so wavevector $\mathbf{q}_A$ decreases, whilst the partnering $\mathbf{q}_B$ increases. Plotted in figures 4a,b of [1] is just how the measured scattering amplitudes alter with wavevector magnitude firstly to specify $|\mathbf{q}_A|$ and $|\mathbf{q}_B|$, and how secondly the latter two peak scattering positions shift progressively in *k* space as the binding energy of the STM tip is incrementally augmented. By combining this information one is able, upon accepting the above dispositions for $\mathbf{q}_A$ and $\mathbf{q}_B$, to mirror the Fermi surface. The outcome was not actually displayed in [1] and is shown now in fig. 1. Included in the figure is a circle about the zone corner that would for an idealized, homogeneous, band-like situation in the 2D limit correspond to F.S.for a hole doping $p_h$ of (1)+ 0.16, the band filling to yield maximum $T_c$ in all cuprate HTSC systems.

**{figure 1}**

The HTSC compounds of course are not ideal metals but very highly correlated, are not homogeneous but chemically mixed-valent and dynamically charge- and spin-striped [2e,d], and are not in the 2D limit but hold measurable 3D character. In the Bi-2212 case the latter automatically will introduce direct intra-bilayer interaction and for a standard material this is going to lead to a two-sheeted $d_{x2-y2}$ symmetry Cu-O *pd*σ* band complex. Here the fuller component (re *electrons*, and the one supplying the outermost electron F.S. sheet) will be the one for which the intra-bilayer interaction entails *c*-axis 'bonding' as opposed to 'antibonding' phasing between the pair of *pd*σ* wavefunctions introduced by the two Cu-O chessboard arrays per unit cell. Following much initial searching for this 3D splitting several more recent ARPES works employing better resolution are able now to detect both components of the *pd*σ* band [10-12], at least for optimally and overdoped material. Because the same groups however when working with Bi-2223 observe only two dominant spectral features, not three [13], and more tellingly yet with Bi-2201 continue to report two clear features, not just one [14], this justifies our continued adherence to the position re-expressed recently by Campuzano, Norman and Randeria [15] in their very comprehensive review of the ARPES work – namely that the universal peak and hump circumstance apparent in all these spectra is primarily the consequence of strong correlation and scattering, and arises quite independently of any multisheet coupling. The consensus to emerge from these more refined

ARPES determinations now as regards the overall shape of the Fermi surface is that the latter departs appreciably from the circular form of the idealized surface present in figure 1 towards a cross, retracted somewhat in the vicinity of the $45^o$ nodal directions and correspondingly inflated in the saddle point regions. It is pleasing then to observe that this is precisely the kind of modification indicated in figure 1 by the new STM data. The curve seen there is very comparable in form to that for the fuller F.S. component in the LDA band structure calculations [16], and likewise to that tracked by the uppermost spectral feature (sharp peak) in the ARPES work. In the ARPES data the two key spectral features ('peak and hump') are separated at the saddle points by ~ 90 - 140 meV, dependent upon doping level. The LDA band structural work [16] produces a bonding/antibonding splitting at the saddles under a coherent bilayer-type interaction of up around 300 meV, but in contrast the new ARPES work would indicate a much smaller value of around 85 meV. By symmetry the latter value steadily becomes reduced to zero on moving towards the $45^o$ direction. It is for very different reasons that the $\pi,\pi$ direction is the orientation to be adopted by the superconducting *d*-wave nodes [2b].

At this juncture it is well to recall that de Haas-van Alphen-type resonances have not yet been reported for the HTSC cuprates, even from that most favourable of cases $YBa_2Cu_4O_8$ (Y-124) [17], and in all probability this will not be attainable either with very highly overdoped material, witnessing the parallel null result obtained for the 3D *s*-electron mixed-valent superconducting system $(Ba/K)BiO_3$ [18]. All HTSC materials are characterized by chronic scattering and this is not just because they are poorly formed or intrinsically nonstoichiometric. Indeed with $PrBa_2Cu_4O_8$, upon compromising greatly the potential there within the Cu-O basal *planes* for metallic conductivity (and indeed superconductivity), this chain-bearing material in fact becomes rendered a notably improved metal, more coherent and Fermi-liquid-like, even in its *c*-axis direction [19]. (Recall in the case of Pr-124 that despite it macroscopically being crystallographically perfect, one finds the basal plane electric and magnetic behaviour to be totally transformed under the strong Pr*4f*-Cu*3d* hybridization. The latter *f-d* mixing with its spin coupling inhibits production of the RVB spin-singlet condition so essential to permitting HTSC to arise.)

Within the author's negative-*U* two-subsystem perception of HTSC [2], the very severe S = 0 pair scattering experienced in HTSC *Y*BCO-124, -123, etc. by the basal carriers near the saddle points and hot spots constitutes the very means to attaining local pair formation in the key shell-filling negative-*U* process. Saddle point scattering is source not just to the superconductivity but also in large measure too to the highly anomalous 'normal' state properties, in particular to the *p*-type sign for the carriers, (via reinforcement of the strong Mott Anderson pseudo-gapping), and to the $T^{-2}$ dependence in their mobility up to very high temperatures [2e, 20]. As has been elaborated upon recently by Hussey [21], the HTSC cuprate systems exhibit large basal *e-e* scattering anisotropies bearing enormous *T*-independent prefactors, and the customary poor-metal *e-ph* resistance saturation is no longer pre-eminent. The above chronic saddle-point scattering was affirmed to add in parallel to the more normal component, in evidence once away from the saddles. As it happens, and as was portrayed in fig 3 of [2b], given the existing geometries of crystal structure and Fermi surface, even the best of carriers, namely those running in the nodal directions, are open to strong *e-e* scattering across into the saddle-point sinks. It is not therefore surprising that the HTSC materials carry significant local character through to such high doping levels, this in spite of a standard LDA band structure analysis ascribing to the $d_{x2-y2}$ $pd\sigma^*$ band a bare width in excess of 4 eV. That width reflects the large degree of *p*/*d* mixing being incurred as the copper *d*-states drop down towards shell closure through the oxygen based *p*-states [22]. The above extremely strong *e-e* scattering actually persists to well out beyond $p_h = 0.3$ [23], *i.e*. to well beyond where the novel superconductivity can any longer be sustained.

**{figure 2}**

In figure 1 the tip energy has been included as running parameter. Indicated too is the point at which the STM scattering signal maximizes. The latter arises well away both from the saddle point and the 45$^o$ nodal directions. In fact the present scattering falls below detection considerably in advance of reaching either such limiting position. With figure 2 the next step is shown of displaying the variation of the above $E_q$ as a function of $\theta$ around the Fermi surface, $\theta$ here being measured about the zone centre anticlockwise from the $\pi$,0 saddle direction. This plot allows direct comparison to be made between the energies $E_q(\theta)$ and $2\Delta(\theta)$, the latter as set approximately by the first sharp peak in the ARPES spectra formed below $T_c$. Angular peak position plots have been published by Mesot *et al* [24] for a variety of under-, optimally and over-doped BSCCO-2212 samples. It is this data which we now incorporate in figure 2, although here reading it as being much more closely related to $2\Delta$ than to $\Delta$ as originally taken in [24]. Both the $E_q(\theta)$ and the $2\Delta(\theta)$ values are somewhat difficult to quantify exactly and call for a certain amount of smoothing of the raw data. Nonetheless several features stand out from the plots given in figure 2: (*i*) $E_q(\theta)$ is effectively a linear function, (*ii*) $2\Delta(\theta)$ as was emphasized by Mesot *et al* [24] is (as extracted) far from being pure $d_{x2-y2}$ in form, (*iii*) despite this $2\Delta(\theta)$ plot bowing towards diminished $|2\Delta|$ versus the idealized sin 2$\theta$ form, it does ultimately turn toward zero as $\theta \rightarrow 45^o$, (*iv*) $E_q(\theta)$ however moves to zero much in advance of $\theta = 45^o$, (*v*) the maximum value of $2\Delta(\theta)$ looks to be reached in the vicinity of the hot spot where the Fermi surface crosses the (0,$\pi$)-($\pi$,0) tie line near $\theta = 12^o$. No Fermi surface exists, note, close to the $\theta = 0^o$ saddle-point direction, in particular for the 2D limit. What now is viewed as being highly significant is that across the central range of $\theta$ the STM data points $E_q(\theta)$ clearly reside, for common $\theta$, at binding energies that are considerably reduced as compared with $2\Delta(\theta)$, identity being gained only at the stage of maximal $2\Delta(\theta)$ reached with the hot spot location [2b]. All such hot spot locations appertaining to the various sample underdopings would appear to line up along the extrapolated $E_q(\theta)$ linear plot. Optimal doping arises much in advance of this extrapolation driving the hot spot location right back to the saddle axis $\theta = 0^o$. A rough examination indicates that the latter circumstance would not be encountered until $p \approx 0.3$. In all cuprate HTSC systems such a doping level happens in fact to be where the last signs of superconductivity are recorded. Within the present modelling this comes because by $p \approx 0.3$ there no longer is retained any well-defined two-subsystem (mixed-valent) character: the materials are reduced to more standard Fermi liquid behaviour, and the high correlation negative-*U* HTSC scenario has become rendered inoperative [2].

**§3. The local pair uncondensed boson mode as scattering object.**

Norman and coworkers have for some time advocated that the peculiar form of the ARPES spectrum below $T_c$, with its peak, dip, hump structure, points to the interaction of the quasiparticle states with some bosonic mode [25]. The natural implication of the presence of such a feature is that it relates critically to the superconductivity if not actually causing it. Not only does the above spectral activity figure at the gap antinodes, but at the gap nodes too there is evidence of a perturbation of simple behaviour, a clear kink being evident in the quasiparticle dispersion curves, this growing more marked below $T_c$. When Johnson *et al* first reported the latter behaviour they attributed it to spin fluctuations [26]. Subsequently it was suggested by Lanzara and coworkers [27], from analogy with what is found in a standard strong-coupling superconductor like Mo [28], that phonons were responsible. The energy is now however too large (~ 60 meV) to be appropriate for phonons in general, and the situation would have to be more akin to some specific optic phonon coupling, such as is present in a CDW/PSD system like 2H-$TaSe_2$ [29]. Although this action might relate in the cuprates to incipient stripe formation, it is not evident why then the kinking would grow sharply at $T_c$ to show a component growing in magnitude as the superconducting order parameter. Despite spin-singlet pseudogapping being strongly in evidence [9], Norman and colleagues would look to have settled for a magnetic attribution to the mode's origin [30,8a], in view largely of its similarity in energy to the much discussed 'magnetic resonance' excited at **Q** = $(\pi,\pi)$ in neutron inelastic spin scattering – for optimally-doped Bi-2212 a quite sharp feature positioned around 43 meV [9]. The alternative perspective already advocated in [2a,b] is that the mode being sensed in the ARPES measurements is associated with local bosonic pairs – pairs arising with the shell closure, charge fluctuation, negative-*U* process. The latter proceeds in line with the energetics laid out in [2h,22], which are it is claimed in [2c] substantiated by the laser pump-probe experiments [31] and the thermomodulation spectroscopy [32] from Stevens *et al* and from Holcomb *et al* respectively. This interpretation subsequently received further support in similar, more wide-ranging optical work by Little *et al* [33], by Kabanov, Demsar *et al* [34], and above all by Li *et al* [35], results examined at length in [2a]. The situation throughout has been perceived as a fragmented two-subsystem one (see fig 4 in [2h]), at low doping often being driven dynamically towards stripes [2e,d]. This highly perturbed geometrical condition (not entirely unlike that in a quasicrystal), when taken in conjunction with a negative-*U* pair resonance stationed in close degeneracy with the Fermi energy, assures that the level of particle-particle scattering will be extreme. A high flux of interchange between fermionic and bosonic states, as well as between boson states which are condensed and those that remain outside the Bose condensate, is inevitable.

Calculations of a negative-*U* Hubbard type consistently have indicated that when, as here, the overall *effective* negative-*U* value (per pair) emerges as being of the order of the bandwidth, the greatest $T_c$ values will at that stage be met with [36]. In the present case $|U_{eff}|$ is indeed, as $W(d_{x2-y2})$, ≈ 3 eV [2a-c]. As listed in [3,4,5], theoretical (homogeneous) negative-*U* boson-fermion equilibrium models recently have been much extended beyond the original work of Micnas, Ranninger and Robaszkiewicz [37] and of Friedberg and Lee [38]. One key matter specifically to be addressed now by de Llano, Solis and coworkers [5b] is that of boson pairs possessing non-zero center-of-momentum, together with individual bosons instantaneously remaining unincorporated into the Bose condensation. The latter pairs, unlike the condensate itself, will of course show considerable dispersion under the electronic/thermal activation. A sizeable population of uncondensed bosons is able to develop in step with the number of condensed bosons (both negative-$U$ pairs and seeded Cooper pairs). Such uncondensed bosons have been suggested earlier in [2a] as being likely source of the extra component of a.c. conductivity discovered to present itself in roughly order parameter form below $T_c$ within the 100 GHz time domain transmission spectroscopy of Corson *et al* [39]. As the system becomes more underdoped the *fractional* population of uncondensed pairs mounts as the negative-$U$ state itself migrates to higher binding energy. This parallels growth in the 2Δ* value witnessed as a pseudogap.

In developing their bosonic mode modelling of the ARPES data a considerable advance recently has been made by Eschrig and Norman [30b] in incorporating correctly into the self-energy analysis of the mode/quasiparticle interaction the effects of the multi-layer coupling. There emerges a clear picture of just how the peak, dip, hump structure develops within both the bonding and the antibonding components to the ARPES signal. It is the uppermost (*i.e.* antibonding) component that it is revealed bears the dominant peak marking fairly closely the maximal value of 2Δ(θ) – just 32 meV in the underdoped ($T_c$ = 65K) BSCCO-2212 sample employed in [30b]. With optimally doped material the above value has grown to 38 meV. Note the latter binding energy still stands some way short however of the 43 meV of the ‘resonance mode’ so to characterize the polarized inelastic neutron spin-flip scattering [9a]. The latter resonance I earlier have associated with complete destruction of a local pair boson. Such identification matches the above numbers, with 38 meV per pair being allotted to an S = 0 separation of the pair of electrons into individual fermions with the additional 5 meV then imparted in the spin-flip to one partner, this reflecting the minimum spin gap magnitude. The individual components of the pair find themselves back now near the hot spots on the two saddles from which originally they were abstracted in the original negative-$U$ pair production process. The above pair elimination necessitates a net momentum transfer to its electrons from each scattering neutron of $(\pi,\pi)$, with $(\pi-\varepsilon,\varepsilon)$ going to the one emerging on the $x$-axis saddle and the balance of $(\varepsilon,\pi-\varepsilon)$ to the one coming onto the $y$-axis saddle (see [2b, fig 3]. For the case of bilayered YBCO and BSCCO what particularly is noteworthy in the neutron scattering process is that it predominantly occurs associated with a large incommensurate $z$-axis momentum transfer which corresponds in direct space to the Cu-O bilayer spacing. Eschrig and Norman [30b] following the lead of Fong, Bourges *et al* [9] interpret the latter detail as evidence of a dominant magnetic aspect to conditions in the material, and view the bonding and antibonding channels evident in these bilayer systems as being associated respectively with ‘even’ and ‘odd’ parity magnetic spin coupling. They pursue this direction despite the sizeable spin gap, with no evidence in optimally or overdoped material of any genuinely magnetic correlation length divergence towards low $T$ [40]. By contrast the negative-$U$ point of view would see the $z$-axis momentum transfer as being required by charge symmetry, as a local pair is undone and its components separated to nearest-neighbour cells. The two electrons, being left spin-parallel by the action of the neutron, are no longer permitted to occupy the same Cu-O coordination unit (as $d^{10}p^{6}$). The process is illustrated in detail in figure 3.

It should be pointed out here that the above *z*-axis 'complication' is in no way essential to the viability of the HTSC mechanism, it relating solely to pair break up. It is simply a symmetry requirement imposed by the crystal structure. When, as with Tl-2201, one has a single layer structure the neutron spin-flip scattering peak centres on the $k_z = 0$ plane [9b]. It remains to be found how this structure will simplify details also of the ARPES spectrum for Tl-2201.

**{figure 3}**

The {π,π} point composite bosons are able by virtue of their negative-*U* standing to acquire an energy which is near-degenerate with the Fermi energy. Through elastic boson-boson collisions these bosons are from this state capable of being rendered condensed **k** = 0 objects of comparable energy, and as such are open in turn to becoming seeding agents towards Cooper pair formation for further Fermi surface quasiparticles in more standard +**k**/-**k** fashion. The addition of such Cooper pairs will drive up somewhat the system-wide superconductive coherence length. The negative-*U* bosons in the above way potentially can introduce an overall level of pairing wherein the triggering zone-edge component can well constitute numerically the minority species.

Besides the two types of condensed composite negative-*U* boson, the derivative Cooper pairs, and any remaining unpaired fermions (present in particular in our current d-wave circumstance), one as well must anticipate, as indicated earlier, heavy undisrupted bosons which have been excited from their condensate through phonon interaction, also clearly in play in HTSC processes. Inelastic neutron scattering experiments bring to light in fact strong modification to the dispersion of certain basal longitudinal optic phonon modes at very short wavelength ($|\mathbf{k}| \geq ½(0,\pi)$) [41]. Such phonons possess the momentum needed to transfer a composite boson into the saddle regions for the (closely degenerate) Fermi surface as a thermally/electronically excited bosonic quasiparticle, outside either the **k** = 0 or π,π condensates. These excited bosons will be of appreciable lifetime since possessing binding energies somewhat less than for the Bogoliubovon pairs at the same locations in *k*-space. Wherever the dispersed excited bosons either would drop below the chemical potential in the superconducting state (as near the saddles) or rise above the Fermi energy for the unpaired quasiparticles (as near the nodal *k*,*k* axis), there the excited composite particles will become unstable and disintegrate. In the above manner we understand how it is possible then to see materialize a bosonic mode having many of the attributes conveyed by the novel STM scattering results introduced in §2. The detected mode departs upwards from the $2\Delta(\theta)$ gap energies in an ever-increasing fashion as one shifts away from the hot spot location toward larger θ. While the condensate coupling holds to a finite $2\Delta(\theta)$ binding right around to the nodes, the excited bosons disintegrate considerably before reaching the $\theta = 45^{o}$ orientation. The excited boson mode stands sharpest for mid-range θ values because there the excited state lifetime is longest, least affected by the high electronic activity proceeding around both saddles and nodes.

**§4.  The condensed boson mode.**

It has at this point to be emphasized that the above dispersed mode is distinct from the bosonic mode concentrated upon by Eschrig and Norman in their ARPES-related work [8a,30]. The latter mode runs virtually undispersed throughout the outer part of the zone. In the present work that mode will be associated directly with the condensate itself of local pair bosons and lies to higher binding immediately below the dispersed mode of *un*condensed pairs. Extrapolation of the line relating to the dispersed mode in figure 2 down into the ($\pi$,0) saddle-point region serves to place the ground state 'bosonic resonance' (labelled $\Omega_{res}$ by Eschrig and Norman) at approaching 60 meV, *i.e*. for this *under*doped sample at a somewhat greater binding than its maximal superconductive gap energy per pair, $2\Delta_o$. Conversely for the *over*doped ARPES sample dealt with in [8a], Eschrig and Norman's self-energy analysis indicates the little dispersed mode in that case to sit just *above* the slightly diminished $2\Delta_o$ level operative there. The more underdoped (*i.e.* ionic) the system is, the farther below the Fermi level the local pair negative-*U* state resides. We will return shortly to this second mode and to its relation to the neutron scattering resonance peak.

With somewhat underdoped 2212-BSCCO it has been observed that 60 meV is in fact an energy echoed in the self-energy driven 'kink' feature in the ARPES-determined quasiparticle dispersion curves readily visible once clear of the saddle direction [26]. There, under the prevailing crystalline and condensate symmetries, the superconductive and bilayer gapping effects are lessened and the state structure simpler. The persistence of this observed band kinking to well above $T_c$ marks a continued presence there of minority negative-*U* local pairs, while the loss at $T_c$ of the Meissner effect, *etc*. registers the extinction at that point of the majority Cooper pair population, and along with this a relinquishing of global phase coherence. Note in the present view the spin pseudogap existing to a hundred degrees and more above $T_c$ is to be ascribed not so much to superconductive inter-pair phase fluctuation as to dynamic RVB spin gapping. As is stated in [2] this RVB gapping is perceived as the means towards preparing the system for the production of local-pair spin-singlets, as well as it being key in their preservation. The absence of any anomalous peaking in the c-axis IR conductance just above the superconducting energy gap within underdoped yet 'well-formed' Y-124 corroborates that the fluctuational behaviour to be seen far beyond $T_c$ is not primarily superconductive in character [42]. Full participation of the bulk of fermions in the seeded Cooper pairing is required before the systems can take up an overall condition brought somewhat more standard in its transport properties. Once below $T_c$ the dramatic fall-off in the chronic saddle-point scattering of local pair creation and excitation brings about very rapid growth in the residual electronic mean free path, as disclosed for example in the thermal Hall data [43] and in the very steep decline in nmr relaxation rate [44].

Just what the nature of the low dispersion $\Omega_{res}$ mode is, and its relationship to the neutron scattering peak, has been most hotly debated [45]. The uniqueness of both matters within the superconducting field suggests a very close link with the local pair condensate itself. This similarly is true as regards the observed strong and very specific coupling into the problem of the basal-plane Cu-O bond-stretching longitudinal optic phonons. Over the outer parts of the zone these particular phonons find themselves severely depressed in energy [41,46], and in a manner quite dissimilar to that for some regular Fermi surface dictated soft-mode CDW/PLD circumstance [47] (see §6). The fact that this coupling, as with the kink in the quasiparticle dispersion above, is observed too in the *k*,*k*,0 direction rules out any connection with stripe phase formation. The virtually dispersionless form of the ARPES-revealed mode as contrasted with the mode revealed in the new STM work is striking. It brings to mind the phenomenon of second sound in liquid $^4$He. Upon looking into the relevant literature it soon is uncovered that the appropriate excitations have long been postulated for strong-coupling superconductors. What is more, their study actually was advanced some years ago to 2D geometry by Belkhir and Randeria [48] in the wake of developments with the cuprates – and was indeed performed within a negative-*U* context.

What was explored in reference [48] is how the collective mode spectrum of a (homogeneous) superconducting system can evolve as one progresses from BCS weak coupling towards the hard core boson régime across the intermediary crossover régime, wherein the cuprates clearly reside. The paper would indicate the existence of a smooth linkage between the Anderson mode established in weakly coupled BCS superconductors and the Bogoliubov sound mode applicable to the extreme local pair limit. Belkhir and Randeria adopt a site-homogeneous negative-*U* Hamiltonian and go forward then within a generalized RPA formalism to examine the crossover régime. They turn specifically to a 2D geometry and also make the important incorporation of screened and unscreened Coulombic interactions, highly relevant for short-range quasiparticle pairing. The outcome is that the collective mode excitations in the crossover region exhibit small but finite dispersion as $\sqrt{q}$. The energy of the mode, while being raised somewhat above the hard-core boson plasma excitation (which would be rather low-lying since governed by $1/m_b \propto t^2/U$), displays very considerable departure from the high-lying Goldstone-like zero-dispersion behaviour of the Anderson mode, a consequence of the increased mass of the bound pairs within the strong-coupling lattice-type modelling. For conditions where $T_c$ maximizes (*i.e.* where $|U|/W \approx 1$ [36, 3]) the mode is able to pick up a dispersion that amounts to a not insignificant fraction of *W*. Simultaneously it is augmented in absolute binding energy to well below the Fermi velocity-dictated plasma energy of the Anderson mode for weak coupling. Under the real *in*homogeneously doped circumstances of the HTSC cuprates one might envisage the above plasma frequency as becoming depressed not just because of the raised value of the effective mass for the individual bosons but additionally by virtue of their relatively low and spatially variable population density.

In probing the dielectric/optical response of a material it is well known that a plasma excitation is best monitored through the electron energy loss (EELS) function $\mathrm{Im}\{-1/(1+\varepsilon(\omega,\mathbf{q}))\}$. It is this same function that one looks to also for information regarding longitudinal optic phonons. HREELS work on a metal demands high surface quality, it in effect monitoring the surface resistivity. The low temperature cleavage of BSCCO affords one excellent opportunity to secure reliable plasma excitation results, a project in fact carried through back in 1992 by Li, Huang and Lieber [49]. Sure enough what was reported was a single strong spectral peak centred at 60 meV, this displaying furthermore a thermal behaviour which clearly associates it with the superconductivity. The above workers in fact related the signal to pair breaking. Notwithstanding this they reported the 60 meV peak to show a $T$ dependence that despite looking to collapse sharply at $T_c$ would if treated as a mean-field BCS feature actually extrapolate to an 'onset' temperature of 150 K. Accordingly it would seem that our boson mode elicits appreciable energy loss only as it becomes heavily damped upon the engagement below $T_c$ of the majority Cooper pair population.

The presence of a multi-sourced pseudogapping up to temperatures of 150 K and beyond is endorsed in a great many types of experiment, not least optical, but the cleanest indicator to date that this temperature range retains some 'superconductive' content is provided by the very recent Nernst effect measurements from Wang *et al* [50]. The Nernst effect involves the production of an electric cross-potential (y-axis) signal as the passage of charge down a (x-axis) thermal gradient is subjected to a large and mutually perpendicular (z-axis) magnetic field – the thermal equivalent of the Hall effect. The signal issues from normal carriers in addition to entities associated with the superconductivity, but whenever of course the latter exist they are dominant. The above paper actually focusses on the signal that can come from the flux vortices once these enter their 'liquid state', depinned from the lattice and able to drift in the thermal gradient. However because the signal Wang *et al* record clearly extends up 20 or more degrees above $T_c$ it is evident that a local pair boson attribution becomes there most appropriate.

**§5. Expressions of the mixed-valent inhomogeneity.**

What additionally is so instructive about these Nernst data [50] is that they supply a direct means of assessing $H_{c2}$ (at which applied field the strong Nernst signal component naturally is taken to zero). Although with most HTSC samples $H_{c2}$ lies well beyond the 30 teslas experimentally available in [50], it is discovered, as a result of there emerging simple scaling rules for both reduced fields and temperatures, that the zero temperature critical field can be extracted as a function of (under-)doping right up to 150 tesla (for $p$ = 0.08). The $H_{c2}(0)$ values so deduced are directly convertible to the corresponding coherence lengths $\xi_o$ { = $(\phi_o/2\pi H_{c2})^{½}$} and manifest the latter as a monotonically rising function of doping $p$ (i.e. of metallicity), $\xi_o$ being only 10 Å down at $p$ ~ 0.05 while close to 20 Å (or $5a_o$) by optimal doping. This reflects that as the participation of Cooper pairs is advanced the extreme local character of the superconductivity becomes relaxed somewhat. Despite $n_s$ growing in proportion to $p$ through this doping range [7], the pairing force itself ultimately peaks, and so accordingly does $T_c$. For some time it has been established from analysis of the electronic specific heat following Loram and coworkers [6] that the overall condensation energy per dopant charge is diminished to either side of a 'critical doping' (just above the optimal for $T_c$), and in particular towards the underdoped side. As was pointed out in [2h], concentrations of hole doping of the parent Mott insulator very close to the optimal level are facilitated in this optimization of properties by virtue of the percolation limit that they represent within the two-subsystem basal plane geometry.

In view of the above it is rather strange Loram, Tallon and Liang in a recent preprint [51] have chosen to play down the fragmented circumstances prevailing in the cuprate systems, especially when underdoped. They declare that the rather sharp aspect to what occurs at critical doping would imply a more uniform state, and they advance their view by reference to sharp signals seen in nmr work specifically for $^{81}Y$ and $^{17}O$. However the type of disorder being addressed in [51] is restricted to that associated solely with static doping disorder and frozen charge segregation, as the recent STM results could at first sight appear to flag [52]. In all HTSC systems (bar $YBa_2Cu_4O_8$ and fully oxygen loaded $YBa_2Cu_3O_7$) there automatically exists the substitutional or interstitial disorder introduced to raise the carrier dopant count to the desired level. At low temperatures this atomic substituent disorder *is* frozen and generally random. However the key question in regard to the metallic and superconducting properties (Y-124 included) is what is the spatial distribution of the coupled hole content per Cu beyond unity. What then is registered becomes a matter of time scale for the particular experimental probe employed. Wherever charge carrier dwell times are long compared with the characteristic probe time one will obtain a broad and multi-sited signal, whereas, and notably in nmr, if local dwell times are (relatively) short some motional narrowing of the signal ensues. It has been demonstrated in recent $^{63}Cu$ zero-field nmr work (from LSCO) [53] just how upon a thorough analysis the spectra gathered clearly uphold a situation wherein fluctuating regions (if not domain walls) of segregated charge indeed arise. It would be most valuable to see this work repeated now for high quality Hg-1201, a system in which the hole doping is secured via interstitial oxygen rather than cationic substitution. The mercury materials, as with all others examined, certainly show evidence of a $^1/_8$ anomaly [54]. Within the fluctuational mêlée the *local* effective doping level has proved in the LSCO case to be such that it can stand far different from the mean value [53], to such a degree that magnetic moments can in fact emerge at select sites within even mildly underdoped material. This is what it has been argued in [2d] leads to the weak spin-flip effects recorded in μSR [55] and neutron diffraction [56]. Indeed it is the Swiss cheese view of the situation, supported at one time by Loram, Tallon and coworkers [57], although now given a dynamic time scale – more like cheese fondue. The percolation threshold is a sharp threshold and clearly one of considerable import for global superconductivity within these strongly coupled systems. The μSR results early on revealed how the averaged $n_s$ count achieved under hole doping $p$ climbs linearly, and with $n_s$ likewise $T_c$, as marked in transport related properties – in the case of μSR via the screening controlled penetration depth [7]. Under thorough analysis the specific heat results [6] disclose how, nonetheless, a fully proportionate condensation energy is not attained in advance of critical doping; only with the latter is proper superconductive coherence between the different entities and structural fragments within the system fully attained. It is in this way that HTSC systems come to acquire glassy characteristics below $T_c$, especially when underdoped [58].

A striking close up affirmation of the local conditions prevailing in the superconducting state of HTSC materials is to be found in the low temperature, high spatial and energy resolution STM mappings of BSCCO samples published in a fairly recent paper from Davis, Lang and coworkers [59]. Let us inspect this remarkable new data from the perspective of the current paper. Spatially one finds a condition highly fragmented as regards energy gap magnitude, and this within a sample of bulk crystallographic near-perfection. The latter categorization neglects (*i*) the dopant excess oxygen, (*ii*) some Bi on Sr sites, and (*iii*) the Bi-O layer supermodulation. These atomic

‘imperfections’ are all frozen in at 4.2 K and their electrostatic potential creates a time-invariant backdrop for the dopant charge. The latter adjusts and fluctuates at the unit cell level to establish the dynamic two-subsystem environment in which the boson-fermion negative-*U* superconductive action proceeds. In the energy-resolved tunnelling one sees some nano-regions to settle into gaps of around 39 meV, with others up around 58 meV, and a bridge of values between the two. In ref.[59] the former locations are termed α-regions and the latter high gap regions β-regions. From what we have argued already we assert that the above regions are to be associated with Cooper pair and with local pair predominance respectively. In line with this identification Lang *et al* discover the following: (*i*) the β-regions diminish in relative *areal* weighting as hole doping progresses – in a lightly *over*doped sample they form just 10% of the field of view as compared with 50% in an underdoped sample of $T_c$ = 79 K; (*ii*) the conductance within the majority carrier, small gap α-regions is appreciably greater than in the local pair β-regions, (*iii*) the local conductance signal amplitude and the coupled maximal gap size accordingly convert in *anti*phase upon STM tip transference from the one type of region to the other; (iv) only the low gap (Cooper pair) α-regions display an impurity (≤ ½% Ni) intragap resonance signal under a *positive* applied voltage (+18 meV), as would come with Bogoliubovon-like hole state behaviour. While these results in general are supportive of our present modelling there is one aspect which by their nature they cannot show – namely any appreciable organization away from random towards stripe geometry, in consequence of the transient aspect to the latter organization under charge hopping, as was evidenced in the Cu nmr work [53]. The intermediate gaps provide nonetheless some mark of this activity. The α and β regions are remarkably sharp-edged though, and, what is more, are remarkably consistent in their interior gap magnitudes, justifying in this our ‘two-subsystem’ terminology.

The above discussion of the mixed-valent inhomogeneity and its consequences for these systems is very much in line with ‘simple’ theoretical modelling attempted already within the context of a negative-*U* Hubbard Hamiltonian. Working with *s*-wave coupling geometry and employing the Bogoliubov-de Gennes procedure, Suvasini, Gyorffy and coworkers obtained interesting initial results [60], and the problem subsequently has been carried through in considerable detail for intermediate coupling (although neglecting Coulomb interactions) by Ghosal, Randeria and Trivedi [61]. One of the latters’ main results is that for the *s*-wave case a global effective order parameter persists, and this moreover in conjunction with a non-vanishing spectral gap. It would clearly be of much value now to know whether conversely for strong local *d*-wave coupling a vanishing or near-vanishing spectral gap might persist, as experiments on the cuprates would seem largely to suggest. Contrasting work within the *t-J* model plus *d*-wave coupling, but adhering to the spinon-holon view of HTSC, also has made investigation of the consequences for the system of the introduction of strong local disorder [62] (in that paper parametrized through the (screened) effective distance of the dopant ions from the active Cu-O plane).

## §6. A role for phonons.

Just published this month by Davis, McElroy and coworkers is yet another STS paper with novel results to bear strongly upon the above [63]. Following on from the Hoffman paper [1], they include now a reconstruction of the F.S. such as was given earlier in figure 1, but secured this time by identifying and simultaneously fitting *all* eight symmetry equivalent elastic scattering vectors at energy $E$ within the Brillouin zone, $\mathbf{q}_1$ to $\mathbf{q}_8$, apposite to the Fermi surface geometry, not just the shortest two detected previously (then labelled $\mathbf{q}_A$ and $\mathbf{q}_B$). This new more detailed work still asserts though an identity between the quasiparticle Bogoliubovon states and the STM conductance registered states. By contrast the latter we shall continue to claim actually lie at appreciably different binding energy to the former (assessed via ARPES). We feel the STM detected states in reality do not display the dispersion and intensity characteristics to match Davis and colleagues' interpretation of events. The experimental mode dispersion curve extracted in [63] remains much like that given in figure 2 above, if anything presenting a slightly counter-sigmoid form, and certainly does not give one any confidence at all that it could swing round to zero binding by $\theta = 45^o$. Perhaps the best argument to support the Bogoliubovon interpretation profferred by McElroy *et al* in [63] is that quite similar results now are revealed to show up at *positive* bias also. We shall return to this latter point in due course.

Following on from the discussion of previous sections it is best however to deal first with the observations appended to the new paper [63] in its closing paragraph. The authors put on record there, along with figure 4, that the direct space conductance mapping signal, $g(\mathbf{r},\omega)$, exhibits under specific local conditions a novel fine-grained 'tweed' structure. The period of this tweed is just *one* unit cell, and it materializes within the map *only* when and wherever the tip tunnelling energy matches the locally relevant superconducting gap *maximum*. Both italicized observations serve to indicate that the tweed structure is associated with the strong scattering activity proceeding at the zone edge. Scattering under a reciprocal lattice vector G signals the occurrence of a coupled umklapp phonon process, and one which arises seemingly regardless of whatever the *local* gap size maximum at the saddles might actually be. In the present model such lattice involvement already has been mentioned in regard to satisfying momentum conservation as local *pairs* (*i*) are shifted between the (00) or $(\pi,\pi)$ negative-$U$ related states and the saddles, or secondly (*ii*) convert at the saddles into +**k**/-**k** Cooper pairs. In both cases the net momentum per pair demanded for these changes is close to (1,0) or (0,1).

Just which phonons might be involved here is to be gleaned from inspection of the phonon dispersion curves obtained via inelastic neutron scattering. As noted already, there has been discovered to occur appreciable phonon softening (vis-à-vis the corresponding Mott insulating material) of certain basal LO modes towards the basal zone edge [41]. Clear intensity changes furthermore are found to develop between branches and also within a branch as a function of temperature. More striking still, there at first sight would seem to be present 'extra' features in the spectrum beyond those branches anticipated from a standard normal mode phonon analysis for the given crystal structure. Additional extensive neutron scattering data clarifying these matters have just been published by Chung et al [64], and they deserve close examination now with regard to phonon coupling, this due not only to the changes above but also to the local pair boson modes discussed earlier.

Chung *et al*'s work [64] deals in fact with $YBa_2Cu_3O_{6.95}$. Because very large crystals are required for the neutron work, the samples of the orthorhombic material were not *a*/*b* detwinned. The endemic twinning has the unfortunate effect of superimposing the TO and LO phonon branches sensed by the neutrons ($I \propto (\mathbf{Q}_n.\boldsymbol{\varepsilon})^2$) for the given propagation and displacement directions. Accordingly mode separation within the experimental data demands careful scrutiny.

**{table1}**

Basal modes associated with phonon propagation down the crystallographic *a* axis take irreducible representations designated by $\Sigma$, while those running in the *b* axis (chain) direction carry the label $\Delta$. The highest energy axial branches ${}^{LO}\Sigma_{1,4}$ and ${}^{TO}\Delta_{2,3}$ issue from a zone centre $\Gamma$ state of point group symmetry $B_{2g}$ at 72 meV, which Raman work detects also. Modes ${}^{LO}\Delta_{1,4}$ and ${}^{TO}\Sigma_{2,3}$ emerge from a $\Gamma$ state of symmetry $B_{3g}$ lying at 66 meV, again seen in Raman spectroscopy. Already this 6 meV splitting, a consequence of the presence of the chains, is quite sizeable, and effects become more marked once well away from the zone centre. The subscripts on $\Sigma$ and $\Delta$ above specify states covered by particular irreducible representations within the axially relevant subgroups. In each axis a couple of phonon states are to be distinguished, the outcome of there being in YBCO-123 two $CuO_2$ planes per unit cell. Because inter-layer coupling is in general weak in the HTSC structures, such pairs of states find their degeneracy lifted noticeably only as there develops additional more interesting electronic coupling into one or other partner. In order to keep track of the situation a close study of table 1 is recommended. The key action is witnessed in the longitudinal modes, as befits charge coupling, and is especially marked for one of the above ${}^{LO}\Sigma_{1/4}$ branches. These particular ${}^{LO}\Sigma$ branches have both their propagation and their polarization vectors aligned perpendicular to the chains (*i.e*. parallel to *a*), while the two complementary $\Delta$ modes, ${}^{LO}\Delta_{1/4}$ , conversely have both vectors set parallel to the chains (*i.e*. parallel to *b*). No odd behaviour is evident within their associated TO branches, whether in the form of any softening or splitting or intensity change as a function of temperature, and the same looks true too for the bond-bending modes at lower energy. At first sight the apical modes look equally uninteresting, but note here that there is present a nearly constant offset right across the zone of some 6-9 meV between the Raman and I.R. active partners $\Sigma_1$ and $\Sigma_4$ (or $\Delta_1$ and $\Delta_4$), with the simpler symmetry $\Gamma_1$-compatible states taking the lower energy. These apical modes in spite of being formally transverse are in fact dominantly bond-stretching in nature, like ${}^{LO}\Sigma_{1,4}$ and ${}^{LO}\Delta_{1,4}$. The latter two sets of *basal* LO branches each exhibit considerable downward dispersion into the body of the zone, but what really marks out their behaviour as being most unusual is their intensity change with wavevector, recorded in [64] figure 5. As against a standard shell model fitting to the phonon data, there is seen to occur an enormous gain in intensity of specific branches in the vicinity of the hot spots near the zone edge, as well as energies shifted down to ~ 55-57 meV (see [64] figures 5 and 6). While both the basal ${}^{LO}\Sigma$ and ${}^{LO}\Delta$ branches each move into this range, the differing manner of their doing so is most revealing. What is recorded is that the effective degeneracy between the $\Sigma_1/\Sigma_4$ interlayer combinations suddenly becomes lifted halfway across the zone, and it remains so right through to $\pi$,0. The outer segment of the $\Sigma_1$ branch would look to have dropped in energy below the ${}^{LO}\Delta_{1,4}$ branches, which possibly remain unsplit. (N.B. there is of course no interaction here between the $\Sigma$ and $\Delta$ branches - they are in mutually perpendicular orientations within each particular twin domain.) The means whereby the depressed segment of the $\Sigma_1$ branch is strongly acquiring intensity is clearly from some further $\Sigma_1$ mode residing in the vicinity of 55 meV – and this is not the bond-bending phonon mode, which stays little changed in intensity right across the zone ([64], fig. 5b). The changes wrought in the intensity of the anomalous $\Sigma_1$ phonon branch segment manifest two very revealing characteristics; (*i*) as known for some time [41] the effects become more pronounced with the hole doping, *p*, and (*ii*) they grow with temperature reduction below $T_c$, actually tracking there the general form of the superconducting order parameter. However, yet more telling is the sudden onset to this strong interaction as the phonon wavelength passes below $5a_o$ or 20 Å. (N.B. in the colour figures 2, 7 and 9 in [64] red denotes an ($\omega$,k) sector showing a gain in intensity upon *cooling* down through $T_c$, whilst blue denotes an intensity loss.)

It becomes necessary to examine then what is the nature of this $\Sigma_1$ mode to which the Cu-O bond-stretching ${}^{LO}\Sigma_1$ phonon branch is coupling so strongly. That there is additionally a clear although less dramatic coupling to ${}^{LO}\Delta_1$ and to the apical bond-stretching phonons $\Sigma_1/\Delta_1$ would implicate direct charge coupling. Our discussions of §3 point immediately to the plasma-like mode of the local pair boson condensate. The latter mode was taken to sit, remember, at binding energies sited just less than for the bosonic local pair condensate ground state, and was deemed as being responsible for the kinking introduced near $\mathbf{k} = \mathbf{k}_F$ at binding energy $\approx$ 55 meV in the ARPES extracted (BSCCO-2212) quasi-particle dispersion curves – a feature that becomes more marked below $T_c$. This energy value for the condensate plasma mode for wavevectors $\approx \mathbf{k}_F$ we have seen to be compatible with the energy of the ($\pi$,$\pi$)-imparting resonant spin-flipping excitation by neutrons of individual pairs which brings about their ejection from the $k$ = 0 superconductive condensate and ultimately return of their components to $E_F$ at the saddles – an energy of 41 meV per pair in $YBCO_7$. Such numbers would imply that the pair plasmon mode actually disperses slightly downwards (*i.e.* to bigger $|\omega|$) from the zone centre towards the zone edge saddle within these d-wave superconductors. (Such behaviour would parallel that of the 41 meV spin resonance excitation itself, which as was discussed at some length in [65] disperses to numerically *smaller* energies upon a reduction from $\pi$,$\pi$ in the crystal momentum input being suffered there.) That comparable effects to the above occur too in the phonon dispersion behaviour of LSCO [41] signifies that in $YBCO_7$ they do *not* enter solely as a consequence of the crystallographic presence of the chains. The standard shell model does not indicate any such a strong divergence as that observed to develop between and within the relevant $\Sigma$ and $\Delta$ modes.

Why the ${}^{LO}\Sigma_1$ phonon mode with its displacement vectors perpendicular to the chains should couple more strongly with the plasmon mode than does ${}^{LO}\Delta_1$, with its chain-paralleling displacements, would seem largely to be a matter of compatible symmetries. The former mode involves, note, the bumping together of the chains, and this should effect carrier passage between the charge stripes, which in YBCO are known to follow the chain direction [66]. More significant is the reason as to why the really strong coupling sets in only as $k$ increases beyond ~ ½($\pi/a_o$), *i.e.* as the wavelength falls below 4$a_o$. The latter distance, recall, is expressly equal to the pair coherence length ξ for optimally doped HTSC cuprates, and as pointed out in [48] it only is for wavelengths inside this limit that the local pair plasmon mode is able properly to form and so to couple to the lattice.

Very short wavelength charge pair plasma oscillations are not confined to the above basal plane states. In the bi-, tri- and higher order layer HTSC compounds there is below $T_c$ at least partially coherent superconductivity in the $c$-axis direction, with pair transfer between adjoining $CuO_2$ layers ≈ 4 Å apart. These cation occupied spaces are without oxygen and are somewhat more ionic than the covalent $CuO_2$ planes, making for lower local dielectric constants and damping. With tri-layer systems and above, Munzar and Cardona [67] very recently have drawn attention to the fact that certain 'out-of-phase' interplane charge oscillations become Raman active (see their figure 1). Pointedly below $T_c$ the appropriate phonon modes are found to acquire very large intensity and often very substantial frequency modifications are met with upon passage through $T_c$ (*e.g.* in Hg-1234 the $A_{1g}$ + $B_{2g}$ ($x',x'$)-polarized phonon excitation peak at 390 cm$^{-1}$ plummets by 50 cm$^{-1}$). Munzar and Cardona demonstrate, in addition, that the action of these interplane plasma oscillations is to produce a Raman scattering efficiency of such magnitude as to clearly be source to the anomalous form of the $A_{1g}$ electronic Raman feature that so long has plagued satisfactory interpretation of this latter type of Raman work [68].

Comparable effects to the above also have been identified in $c$-axis I.R. results [69], where once below 150 K a broad peaking is observed to develop in the c-axis conductivity centred about 50 meV, with neighbouring phonon peaks at the same time diminishing in intensity. Note through all the myriad means of investigating the small energy excitation range up to 1000 cm$^{-1}$ (120 meV) careful distinction needs to be made between those effects associated with the spin pseudogapping in the DOS (which is not particularly temperature dependent and sensed for example in nmr Knight shift experiments) and those lower energy effects around 400 cm$^{-1}$ (or 50meV) that develop not too far above $T_c$ associated with sharp reduction there in the chronic basal scattering to afflict the fermionic quasiparticles (as sensed in for example nmr relaxation rate experiments). Both types of electronic change scale with $T_c(p)$ [70], though for quite different reasons. The former change relates largely to singlet pair breaking or rather its supression under RVB creation, while the latter change relates to pair creation initiated by the hybrid growth in coupling between fermions and negative-$U$ bosons.

Because the pair plasma mode is interacting so directly with the lattice vibrations, and strongly once below $T_c$, it is not surprising that oxygen isotope effects have extensively been recorded within HTSC systems [71]. The latter effects are however quite different in form from those associated with the standard, phonon mediated, retarded interaction of the BCS mechanism. Indeed it is observed under $p$ change within any given HTSC system that the isotopic shift in regard to $T_c$ itself is pointedly dropping to zero at the very composition to yield the highest $T_c$ – or, more probably, the highest condensation energy per pair and shortest coherence length, the critical doping $p \approx 0.19$. Very significantly, however, there is in operation at precisely this juncture a strong isotopic mass exponent below $T_c$ with regard to the penetration depth $\lambda$ ($\propto n_s/m^*_b$). Because $n_s$ is not appreciably affected by isotopic substitution this must mean an isotopic shift is operative deriving from the bosonic effective mass $m^*_b$. The *electronic* masses are responding here, via the plasma mode coupling discussed above, to the eigen-energy changes being isotopically imposed upon the phonon modes. On cooling from 300 to 4 K, $m^*_e$ itself has been evaluated to mount steadily from 2 to 4 $m_e$ [69].

Since the kinking in the quasiparticle dispersion curves and likewise the development of the peak, dip, hump structuring to the ARPES energy spectrum, each become more pronounced with an increase in layering number, *n*, across a sequence such as Bi-2201, Bi-2212, Bi-2223 [72], one would now much like to find out what happens to the above penetration depth isotope effect. The most probable scenario is that these layering effects actually issue from the known reduction with *n* in the *basal* plane lattice parameter, the latter rendering the negative-*U*, closed-shell state more stabilized, but without any rise here in pair breaking as would arise through underdoping. It correspondingly is registered that uniaxial pressure applied *within* the basal plane produces a steep growth in $T_c(p)^{max}$ [73], and so potentially directly in $n_s$.

As is to be expected from the current modelling, the mass enhancements arising in the saddle regions should be greater than those encountered within the nodal regions of the Fermi surface. This plus a sharp growth in renormalization of the imaginary part of the self energy below $T_c$ under the action of the boson mode is very clearly to be found in the new ARPES data published very recently by Kim *et al* [74] (scanning from $\pi$,0 across $\mathbf{k}_F$ towards $\pi,\pi$). Apparent too in this new data is that no fundamental difference of response exists between the bonding and antibonding interlayer coupled sub-bands. The kinking is more evident in the former simply because it is the fuller. Above $T_c$ the antinodal coupling constant is found to become rapidly much decreased (although still around 1), whilst the mass (but not the scattering rate) is rendered more or less isotropic. Note the quasiparticle mass renormalization resulting from the self-consistent action of the electron pair mode manifests quite different thermal characteristics to what would be expected if the mode were primarily phononic in nature or one of spin fluctuations.

**§7. More on the nature of the HTSC mechanism.**

In the present form of modelling what in large measure dictates the sensitivity of response to parameter changes like pressure and doping is the precise location of the double-loading negative-*U* fluctuational state in relation to the Fermi energy – or once below $T_c$ to the chemical potential. Remember the small energy ~ -50 meV by which the state is to sit below $E_F$ is but a tiny fraction of $|U_{eff}|$, which in the present case is ≅ 3 eV (per pair) [2a,c]. The latter in turn is less than half the actual state adjustment energy accompanying double-loading shell closure, since around 4 eV is negated in the requirement to 'back off' the Coulomb repulsion energy. Accordingly a high sensitivity of $T_c$ to all parameter changes is to be anticipated if these remarkable effects indeed do establish themselves as one reaches close degeneracy between the negative-*U* state and $E_F$, and a ready interconversion between charge fermions and bosons.

This degenerate condition very recently has been referred to by Domanski [75] as a 'Feshbach resonance'. It is demonstrated formally in [75] that when the boson and fermion subsystems are disentangled (via application of a continuous canonical transformation) the residual fermion-fermion quasiparticle interaction itself presents highly resonant structure above $T_c$, which for $T < T_c$ becomes reduced to that of the superconducting gap. A further key result for our purposes obtained in [75] is that the overall boson-fermion system foregoes the particle-hole symmetry which characterizes the standard Bogoliubov analysis of the simpler BCS mechanism.

Domanski is not the only one to investigate this symmetry breakage between the hole and electron Bogoliubov-related pair states in the superconducting phase, the matter being extensively examined also by Batle *et al* in reference [5a]. One consequence of the shifting away in the Boson-Fermion model from the 2*e*-2*h* symmetric BCS limit and the mean field approximation, inherent in allowing bosonic pairings to exist and propagate independently embedded in the Fermi sea, is that uncondensed boson modes occur. These possess a novel dispersion character in that their energy varies linearly with **K**, the centre of mass momentum for the pairs. (When Cooper pairs move in vacuum, as in a simple BCS treatment, their dispersion is quadratic). The velocity for this uncondensed mode of bosonic pairs is comparable in size to the Fermi velocity of the unpaired quasiparticles. It is these linearly dispersed modes of uncondensed pairs which it has been claimed in §1-3 of the present paper are being sensed in the energy resolved STM conductance scattering experiments from Hoffman *et al* [1] and now McElroy *et al* [63]. Because the 2D Fermi wavevector $\mathbf{k}_F$ is virtually circular about $(\pi,\pi)$, a plane incident near-perpendicularly to the energy cylinder at $\theta = 22½^o$ bearing linear dispersion from the saddle will intersect the cylinder in approximately a straight line, just as the STM results appearing in our figure 2 (or figure 3c of [63]) would indicate. The above mode is of appreciably greater velocity than for the plasma mode of the condensed bosons addressed subsequently in §§4-6. The velocity of the uncondensed mode being proportional to the pair coupling strength is automatically quite substantial in the HTSC materials. Because the pairing in these d-wave systems is dominated by action at the extensive saddles in the band structure, the interaction parameter $\lambda = N(E_F).V$ becomes considerable even if $V_{eff}$ is itself not enormous. Batle *et al* indeed find that $T_c$ values of 100 K may be attained with $\lambda$ values of only ~ ½ even when employing energy-shell coupling ranges limited to $\hbar\omega_D$ (*i.e.* ~ 50 meV, with $E_F$ ~ 1 eV). These authors in ref. [5a] as a result loop their argument around in favour of phononically driven coupling, in line with the orientation adopted by Lanzara *et al* [27]. In that case, though, there would then be no compelling reason ever to depart from the Bogoliubov *e-h* symmetry of the traditional mechanism.

Batle *et al* [5a] cloud the situation somewhat further by noting, after Hirsch [76], that the great majority of known superconductors have *p*-type Hall coefficients, as against most non-superconducting metals being *n*-type in their transport properties. This carrier electron-hole asymmetry of course is far from synonymous though with the BCS related usage of those terms. The present author already has commented upon how the observed preponderance of *p*-type materials possessing elevated $T_c$ values appears ascribable to chemical bonding effects, even within the simplest of systems. Where such effects become particularly evident is with the superconducting, high pressure, semimetallic, homopolar-bonded forms of elements like B, Si, P, S and I [77], and seen even more recently in Li [78]. One might include here cluster structures like the $C_{60}$ [79] and $Ga_{84}$ [80] derivatives and the graphite-sulphur composites [81]. The most striking example of bonding-driven pairing, outside the closed-shell effects of the cuprates and bismuthates, is the case of $MgB_2$, where the holes in the σ-bonding B-B band clearly are responsible. As with the cuprates, the fact that the lattice responds strongly to the pairing interaction – here in the unique reaction of the $E_{2g}$ bond-breathing mode – does not mean one ought to invert the terminology and revert to an electron-phonon ascription. Those interested further in $MgB_2$ are directed to the work of Pietronero and colleagues [82] and of Mazin and Antropov [83] dealing with these band Jahn-Teller type effects and with the non-adiabatic, anharmonic lattice coupling involved. Note when talking about 'chemically' driven superconductivity one should take care to distinguish between the above homopolar bonded systems (of which β-HfNCl organo-solvated by $Li^+$ ($T_c$ = 25½ K) constitutes a further recent case [84]) and those reliant upon closed shells and disorder-forestalled disproportionation, such as the cuprates and bismuthates. A very recent additional example of the latter type could be $Na_xCoO_{2+\delta}$ (reported $T_c$ of 31 K [85]). This latter instance would revolve around the high stability of the low-spin $t_{2g}{}^6$ closed subshell configuration, as was intimated in § 7.2.9/10 of [2h'].

Let us nowreturn to the question of 2*e*-2*h* symmetry in the Bogoliubov sense - or, rather, the lack of it. The breaking of such symmetry is a feature of the Boson-Fermion modelling, and one may question whether it finds clear cut experimental expression anywhere. Actually what has long been known is that pair tunnelling signals in HTSC systems are not symmetric between positive and negative bias, unlike for standard BCS superconductors. Domanski and Ranninger have indeed addressed just such matters regarding tunnelling in one of a series of recent papers to employ the B-F model [4c]. Despite formal criticism of their theoretical procedures by Alexandrov [86], the general level of progress being encountered in this approach suggests that the technical problems pointed to in [86], of singular divergences and cancellations, can probably be removed by making the modelling somewhat more complex, as say by incorporating the action of the spatial inhomogeneities of doping and stripes present in the real systems. The participation of uncondensed as well as condensed bosons would appear well-established from the gigahertz a.c. conductivity measurements of Corson *et al* [39;see 2a]. That matter Tan and Levin [87] now have taken forward formally to account not only for the Corson results, but also those from Wang *et al* [50] concerning the Nernst effect in BSCCO as suggested above in §4. Formally they ascertain how up to temperatures considerably higher than $T_c$ a rough mimicking of the Kosterlitz-Thouless behaviour in the former case and a persistence of the Nernst signal in the latter case uphold strongly the presence there of some local pairing. As was indicated already, what fluctuates here, in underdoped systems in particular, is principally the superconductive phase angle between poorly communicating tight pairs, although not quite for the reasons proposed by Emery and coworkers [88]. In this connection Tan and Levin make evident that the transverse thermoelectric coefficient relevant to the Nernst effect provides a much more sensitive probe of the local electronic pairing activity than say the diamagnetic fluctuations. The latter remain not dissimilar in degree to the Aslamazov-Larkin behaviour within a standard superconductor. Ranninger and Tripodi very recently have extended the theoretical modelling of the B-F scenario in this direction [89], now permitting a degree of itineracy to their hard-core bosons: they track then the decay of coupling between the subsystems as a function of underdoping with the shift away from amplitude towards phase fluctuations.

If the above discussed excited bosonic 2*e* pair modes exist for all $T < T^\dagger$ (where the pair fluctuation crossover temperature $T^\dagger$ is, note, not synonymous either with the RVB spin gap temperature or with the charge pseudogap temperature), there ought to be some positive indication of the corresponding independent 2*h* modes. It is our belief that the McElroy STM paper [63] in fact might contain just such an observation, a matter we flagged for subsequent comment at the end of the first paragraph in §6. McElroy *et al* state in [63] that their + and - 14 meV scattering results are only ‘roughly identical’ (see figure 2H versus 2F). Because of the diffuse nature of those results it would not appear feasible as yet to affirm this anticipated difference between the positive and negative bias results without appreciably more care being taken. Nonetheless the general level of difference found between the ordinary $\pm V$ pair tunnelling results is encouraging. It would be extremely interesting now to attempt to follow the STM scattering work to higher temperatures, if manageable without surface contamination.

One thing very apparent through all this work is that the reason the HTSC problem has been with us for so long is that it is a truly complex problem. The previously simplest of experiments repeatedly debouche into a remarkable pan of intricacy which it may or may not be profitable to pursue. The nmr results were an early marker of this [53,90] and the ARPES data more recently have witnessed a similar transition from being treated as beguilingly straightforward to being recognized as highly complicated in nature [91]. What can be said however, as with the evolution of this paper, is that, wherever one probes, the boson-fermion, two-subsystem, negative-*U* approach appears able to cope with more problems than it generates. It is sufficiently complicated of itself to hold the required degree of flexibility to pursue all the various twists and turns that the experimental work exposes. It is self-evident neither the phonon scenario of HTSC nor the spin-fluctuation one possess the necessary degree of built-in complexity to do this. I would refer those still thinking along the latter lines to examine the ‘inversion’ of the photoemission spectra recently presented by Verga, Knigavko and Marsiglio [92], as they seek to account for the observed kinking in the quasiparticle dispersion curves. The spectral form for $\alpha^2.F(\omega)$ of the mode coupling extracted there within an Eliashberg type treatment (see figure 6) is quite unlike what is expected for spin fluctuations, and it in addition contains far too much weight at high energies to be relevant to phonons alone.

One of the chief reasons for seeing the spin-fluctuation scenario actively supported for so long is its perceived compliance with neutron scattering data. Wherever the latter though has been dominated by data coming from LSCO (in consequence of the large crystals available) that is unfortunate since LSCO is the most ionic of all HTSC systems, consequent upon its particular counter-ions [2f]. The RVB spin gap is especially small there ($\approx$ 10 meV), and the spin gap temperature actually looks to fall somewhat below the likewise small $T_c$ [2b]. This has meant many of the observations made on cooling LSCO down below $T_c$ have been regarded as appertaining to the superconductivity when in reality they should have been attributed to the RVB spin gap. The gap studied by Lake *et al* in [93], note, is dispersionless. Additionally down at helium temperatures the spin scattering peaks that the latter register at $E > \Delta_{sp}$ develop to define more sharply a 'spin coherence length' which again is independent of **q** and close to 32 Å or $8a_o$. The latter figure is of course just about the size of the domain in the stripe structure that is being settled into with the opening of the spin gap, as depicted in [2d, fig 1]. The observed cluster of four incommensurate inelastic scattering 'satellites' comes strongly to decorate the *k*-space point $(\pi,\pi)$ in consequence of the dominant 4-spin RVB square plaquet correlations. As was emphasized in [2d] the approximately $8a_o$ size of the incipient domain structure is not Fermi surface governed but simply imposed by the numerology of the doping charge concentration.

Millis and Drew [94] recently have drawn attention to the seeming incompatibility between the self-energy broadened MDC peaks observed in the new high resolution photoemission work as against the known optically determined conductivity. The latter conductivity is considerably better than what in customary circumstances the former widths would match, a dichotomy unexpected if events were to be covered by the spin fluctuation scenario. However, as we have seen, the strong resonant (i.e. quasielastic) scattering activity operative on the $k_F$ saddles associated with fermion-to-boson inter-conversion will not contribute towards restricting the optical conductivity. The large quasi-elastic scattering term in evidence has been commented upon earlier by Abrahams and Varma [95]. They deemed that out-of-plane 'impurities' had to be responsible – presumably the dopant ions. However our view of events is even more 'intrinsic' yet than this. Besides the above sizeable elastic (and strongly angular dependent) term, there exists too of course the very substantial inelastic scattering activity between carriers and between carriers and the lattice. These effects go forward to yield those strong and very characteristic resistivity terms, linear in both $T$ and $\omega$ [96], that Varma early on referred to in his Marginal Fermi Liquid formulation of HTSC behaviour as issuing from a 'magic polarizability' [97]. This carrier scattering is at least an order of magnitude more intense than would occur if the transport properties were simply as covered by a standard Boltzmann treatment and the given LDA band structure. It is not just the unusual $T$ and $\omega$ dependencies to the observed transport properties which are in question, but, above all, the sourcing of their strikingly large prefactors. The situation for the d.c. conductivity has long been recognized, but recently Varma and Abrahams [98] have offered an extensive formal treatment within the framework of MFL of how both the Hall and magneto-resistance signals likewise emerge as being dominated by the novel scattering. Through a small angle scattering treatment they try to attempt to secure the well known HTSC results cot $\theta_H$ ($\propto \mu_H^{-1}$) $\propto T^2$ and $\Delta\rho_B \propto T^{-4}$. However, as they now acknowledge, this is not the right way to reach these relations [99]. It was stressed in [2e] that the latter are not *in toto* Fermiology results; they constitute indeed a clear expression of that intense fermion-boson activity which precipitates HTSC within the present resonant negative-$U$ scenario. Clearly the two-particle, two-subsystem, umklapp and DOS pseudogap aspects to the problem all will require to be incorporated explicitly.

Let us now take stock of the present theoretical situation. The magic polarizability looked to within the MFL theory, with its rather momentum independent and structureless form both energetically and thermally, is in accord with a resonant interconversion/quantum critical fluctuation as comes in the negative-*U* scenario. The action we have seen extends to the lattice, with the accommodation of momentum conservation as quasiparticles transfer between the saddles, the zone corners and the zone centre. The appropriate zone edge phonons are observed to couple into the various charge processes. The local pair bosons still are present to some degree above $T_c$ and they form dispersed propagating modes within the pseudogap. Below $T_c$ the local bosons do not pass in their entirety into the **K** = 0 condensate and excited dispersed modes for 2*e* and 2*h* pairings exist, in addition to the excited sound-wave-like plasma excitation of the condensate itself. The condensate comes to hold as many pairs as the dopant count, as a more standard Cooper pairing around the Fermi surface is precipitated, although the effective overall pair coherence length is kept very small (with $H_{c2}$ correspondingly high). Because of the boson-fermion degeneracy and interconversion, the condensate interacts very strongly with the quasi-particle dispersion curves to create kinks there in all directions slightly below $E_F$. The pairing action itself proceeds by virtual excitation from the band structural saddles to the top of the band at the zone corner, a state that when empty is of high energy under antibonding/bonding correlations. The doubly loaded state however becomes greatly relaxed in negative-*U* fashion by reason of its association with band closure and the complete restructuring locally of all the valence band energy states. The latter comes in response to a termination of *p*/*d* bonding/antibonding operativity referring to the principal Cu-O binding – the essential chemistry in the HTSC problem [2b,2h]. It is a matter of chance that with these particular square-planar cuprate materials the $^{10}Cu_{III}^{2-}$ doubly-loaded state happens to fall in close resonance with $E_F$. Such appears not to be the case for the 3D metallic mixed-valent system $La_4BaCu_5O_{13+\delta}$. The given crystal structure of the HTSC cuprates is doubly favourable in that it introduces a strong zone edge saddle point very close to $E_F$. From the latter it becomes easy to draw off considerable numbers of bosonic pairings of the general type $e(0,\pi) + e(\pi,0) \rightarrow b(\pi,\pi)$, and this without any participation of phonons or spin fluctuations. Phonon participation comes in the main as bosons return to the saddles under the negative-*U* 'relaxation', or in turn pass from the saddles over into the **K** = 0 condensate. The phonons involved in these charge coupled processes are longitudinal optic in nature. Likewise the bosonic mode excitations as plasma oscillations couple to the LO branches, and so become registered via $\mathrm{Im}(-1/\varepsilon)$ in the HREELS experiment.

The 40 meV 'resonance peak' recorded in the neutron scattering experiments [9] is not a bosonic mode capable of supplying the 'glue' for HTSC. Indeed it marks the disruptive excitation of the S = 0 superconducting pairs, involving S = 1 spin flipping, as a pair is taken back from **K** = (0,0) via (π,π) to its component fermions near (π,0) and (0,π). The observed *z*-axis component to this excitation in YBCO-123 and BSCCO-2212 is to do not with magnetic interlayer spin coupling but with the appropriate accommodation of the two dissociating fermionic spins within the bilayer unit cell. The paper by Abanov, Chubukov and colleagues [45b], in preprint form entitled 'What the (π,π) resonance peak can do' [100], acknowledged the spin-exciton view of events endorsed there to be quite distinct from that of antiferromagnetic magnon mediation. Nevertheless they continued still to look towards the action of Fermi sea spanning to build up the response function $\chi'(\mathbf{q},\omega)$ around $\mathbf{q} = \mathbf{Q} = (\pi,\pi)$ and $\omega(\mathbf{Q}) = \Omega_{res}$. This 'across the body of the zone' physics must be contrasted with our 'to the corner of the zone' physics. In order to acquire the large requisite 'spin-fermion' coupling constant, *g,* and dimensionless fermion self-energy coupling constant, $\lambda$, of roughly ¾ eV and 2 respectively it was demonstrated necessary that the 'magnetic' coherence length be just $2a_o$. Accordingly the bosonic coupling mode looked to within [45b] is of very highly overdamped spin excitations, and in this note contrary to what popularly is often advocated – an antiferromagnetic spin fluctuation mode acting as bosonic mediator in BCS-related fashion. Pointedly the above $\xi$ value for the spin coherence length of just $2a_o$ means it is identical to the range of the superconductive coupling in the local pair view [2b,101], a result that exposes the magnetic attribution of the coupling as not being unique. It is self-evident $g$ and $\lambda$ have to be quite large in order to yield HTSC, but the neutron peak has rather little spectral weight or role in events. Clearly the 'hot spots' on the Fermi surface dominate scattering activity in the zone and dictate the novel transport behaviour above $T_c$, besides the *d*-wave form to the superconductive gapping below, but this action patently is not due to Fermi surface nesting. The uncondensed boson modes dispersing upwards from the hot spots, which the energy resolved STM scattering experiments from Davis and coworkers [1,63] now look to bear witness to, reassert the local pair character to the novel events current in these materials – as above all, of course, does the extreme smallness of $\xi_o^{sc}$.

It would appear to the author essential now to try to draw together the many disparate theoretical approaches pursued in the past, including the spin fluctuation and MFL ones, around the negative-*U* and boson-fermion modelling developed in some detail by de Llano, Tolmachev and colleagues [5,102], Tchernyshyov and Ren [103], Letz and Gooding [104], Domanski, Ranninger, Romani, Tripodi and coworkers [4,89,105], and several others. In accomplishing this it will be decisive to embrace properly the inhomogeneous two-subsystem nature of the mixed-valent cuprates, as the STM results of [1,63] and related papers serve so clearly to illustrate.

**Acknowledgments:** I would like to dedicate this paper to Prof B L Gyorffy on the occasion of his 65th birthday and for his helping to make my time in Bristol a long, lively and productive one. At this moment of my own retirement I thank too my wife (P.A.W.) for her inestimable help over a still longer period in seeing this paper and its predecessors brought to fruition. Additionally thanks are due to Dr N E Hussey for his comments on the current manuscript.

**Table 1.**

Characteristics of zone boundary phonons in $YBa_2Cu_3O_7$ propagating *in* basal plane perpendicular and parallel to the chains; *i.e.* along the *a* axis (irreducible representations Σ) and *b* axis (irreducible representations Δ) respectively. Information drawn from Chung *et al* [64], figs. 1 and 4 and text. (Note small errors exist there in figure 1, with 1e being labelled LO rather than TO, and superfluous oxygen atoms appearing in 1c, 1e and 1f at the chain level).

| figure label [64] | in plane propag$^n$ vector (re chain) | displacements (re propg$^n$ vector : LO vs. TO) | displacements or 'polarization' re chain | dominant vibration type re Cu-O bonds | phonon mode irreducible representation | Γ pt. energy (meV) |
|---|---|---|---|---|---|---|
| 1a | x (⊥) | x (// : LO) | ⊥ [Raman $B_{2g}$] | planar stretching | $\Sigma_{1,4}$ | 72 |
| 1b | y (//) | y (// : LO) | // [Raman $B_{3g}$] | planar & chain stretching | $\Delta_{1,4}$ | 66 |
| 1c' | y (//) | x (⊥ : TO) | ⊥ | planar stretching | $\Delta_{2,3}$ | 72 |
| c'' | x (⊥) | y (// : TO) | // | planar stretching | $\Sigma_{2,3}$ | 66 |
| 1d' | y (//) | y (// : LO) | // [Raman inactive] | chain stretching | Δ | * |
| d'' | x (⊥) | y (⊥ : TO) | // [Raman inactive] | chain stretching | Σ | * |
| 1e' | x (⊥) | z (⊥ : TO) | ⊥ [Raman active] | apical stretching | $\Sigma_1,\Delta_1$ | 62.5 |
| e'' | y (//) | z (⊥ : TO) | ⊥ [IR active] | apical stretching | $\Sigma_4,\Delta_4$ | 69 |
| 1f' | x (⊥) | x (// : LO) | ⊥ | bending of both chain & plane. | Σ | 43 |
| f'' | y (//) | x (// : TO) | ⊥ | Ditto | Δ | * |

LO : longitudinal optic, *i.e.* displacement vector // propagation vector, with Cu and O displacements in bonds in antiphase.

Σ vs Δ splitting is from presence of chains destroying tetragonal symmetry.

- the Σ representation wavevectors are //*a* (*i.e.* ⊥ to chains), while the Δ are //*b* (*i.e.* // to chains).

$\Sigma_1$ and $\Sigma_4$ relate to there being two basal planes per YBCO unit cell. (Likewise $\Delta_2$ and $\Delta_3$).

- of which only ungerade (antisymmetric) inter-layer combination generally seen for integer $l$.

## Figures

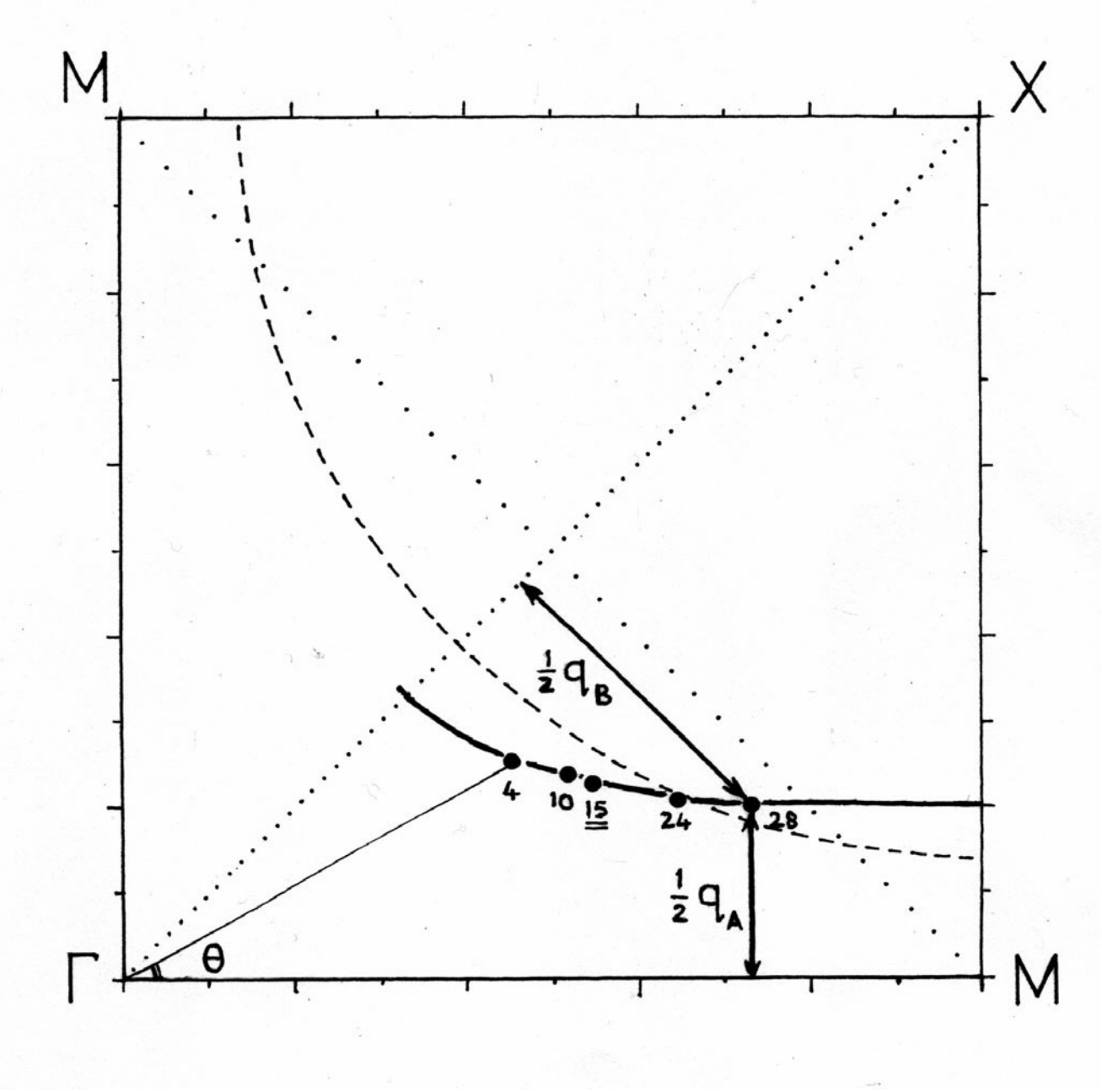

**Figure 1.** Shown in a quadrant of the Brillouin zone is the locus defined by the coupled elastic scattering wavevectors $\mathbf{q}_A$ and $\mathbf{q}_B$ (see text) extracted from Fourier analysis of the STM conductance real space mapping secured at 4 K by Hoffman *et al* [1] for the case of slightly underdoped ($T_c$ = 78 K) $Bi_2Sr_2CaCu_2O_{8+\delta}$. Also shown is the hypothetical circular 2D Fermi surface that would correspond to a hole count relative to half filling (the dashed line) of 0.16, as for optimaly doped HTSC material. The experimental locus obtained here is very similar to that generated in LDA band structural work for the bilayer structured material at optimal doping, as it relates to the fuller of the two sheets to derive from the two Cu-O planes per unit cell (i.e. the bonding interlayer combination). The 'hot spots' are where the M-M dotted line intersects the Fermi surface on the band structural saddles near the M points ('π,0', *etc*.), and the superconducting $d_{x2-y2}$ node lies on the line from 0,0 to 'π,π', close here to one third of the way from zone centre to zone corner. Angular θ values from the $k_x$-axis read off from this plot for each data point are subsequently used in constructing figure 2.

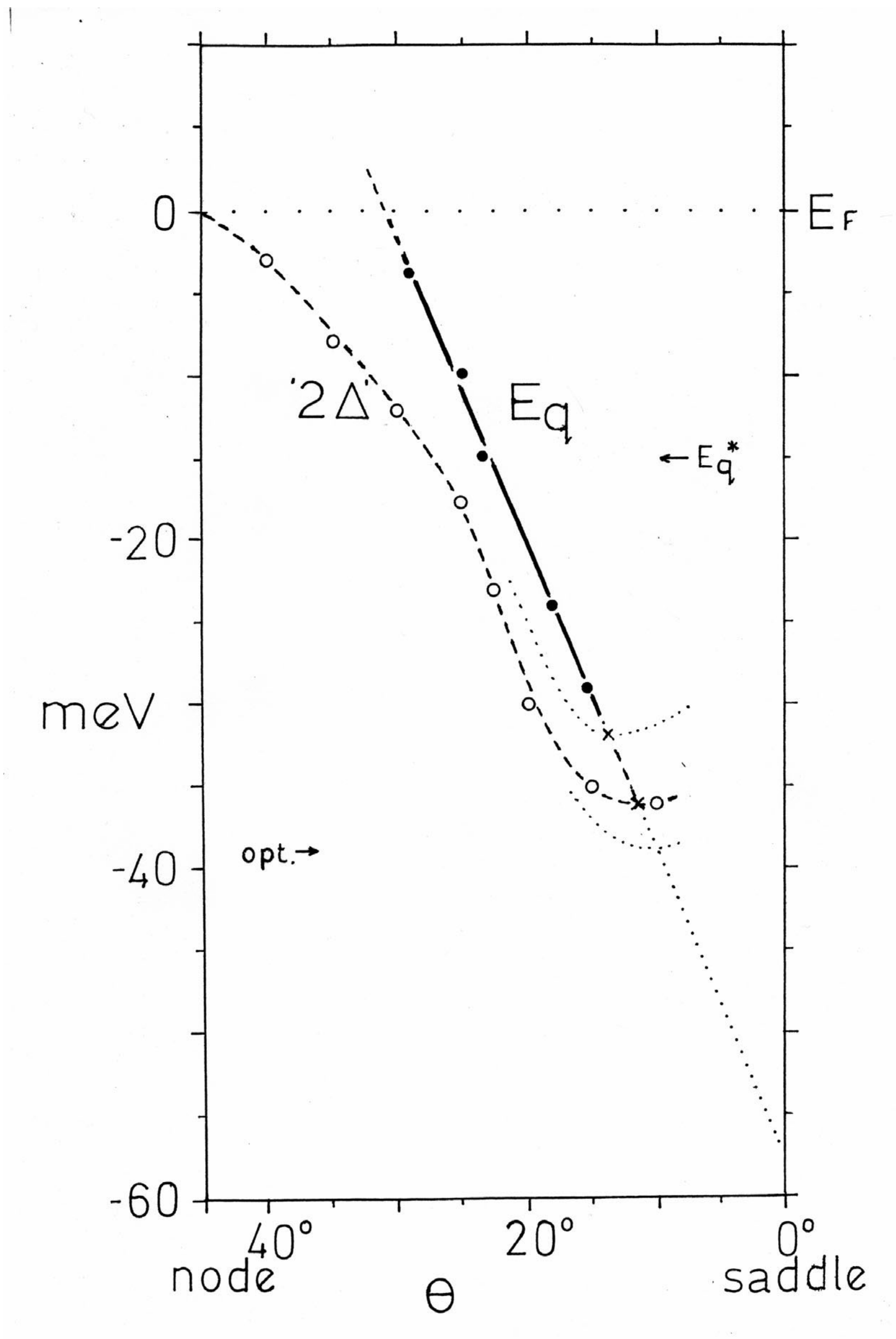

**Figure 2.** The angular variation within the Brillouin zone octant from the nodal to the saddle directions is plotted for the binding energies $E_{\mathbf{q}}$ (at 4 K) corresponding to the conductance data points located at coupled wavevectors $\mathbf{q}_A$ and $\mathbf{q}_B$ in figure 1. The more or less linear dispersion presented is that for a somewhat underdoped sample. The line appears to shift slowly upwards with a rise in sample hole doping content. The diffuse STM scattering signal is steadily lost as on the one hand the experimental conditions approach the Fermi energy and on the other hand are pushed back down towards the 'hot spots' on the saddles. In the present case the sharpest scattering peak signal was registered for the intermediate condition $\theta \approx 23^o$. The above angular behaviour is to be contrasted with that included in the figure of the first sharp peak in the ARPES data under comparable conditions, this constructed from the data of Mesot *et al* [24]. The latter curved plot (dashed) is taken to be a moderately close guide to the functional form of the superconducting gap $2\Delta(\theta)$ – not $\Delta(\theta)$ as stated in [24]. It is not justified, moreover, to take directly the non-sinusoidal form of this plot as representing faithfully any true deviation from a strictly $d_{x2-y2}$ form for the superconducting order parameter.

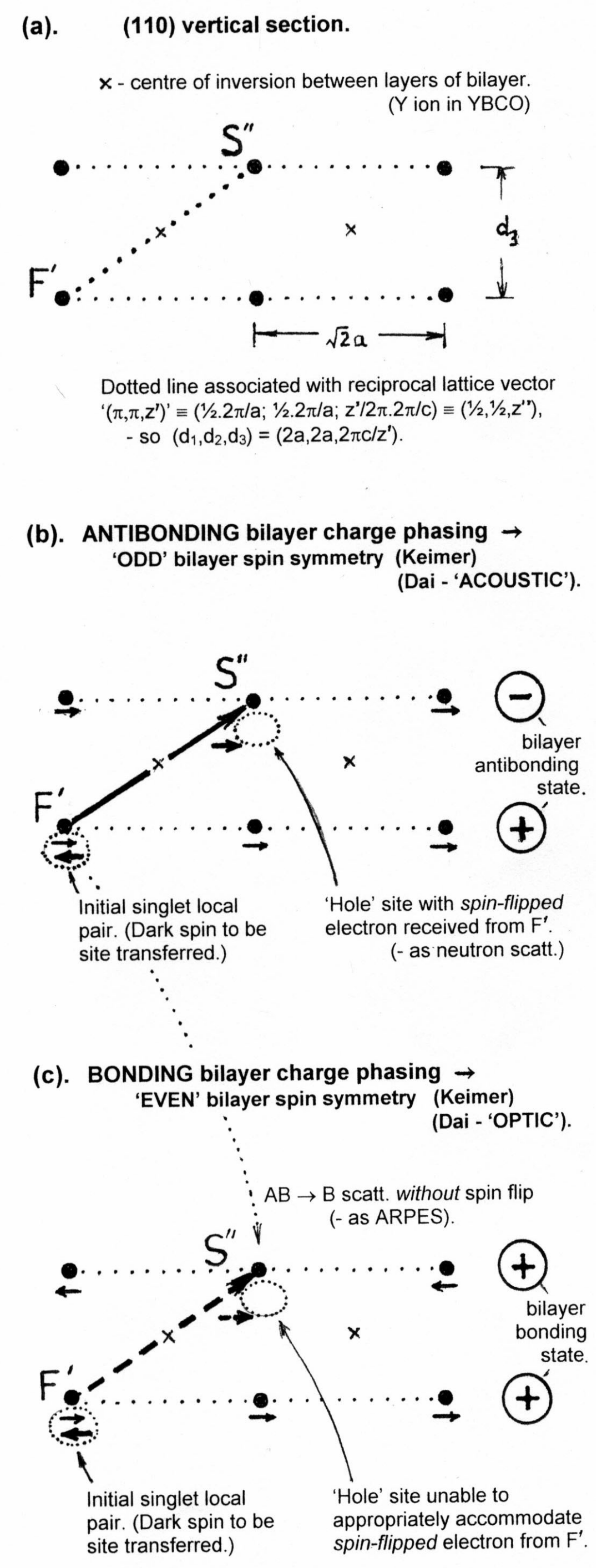

**Figure 3.** The symmetry conditions governing the way in which the bosonic pairs decompose under inelastic neutron spin-flip scattering within the bilayer structures of $YBa_2Cu_3O_7$ and $Bi_2Sr_2CaCu_2O_8$. The situation is shown in vertical (110) section. The appropriate process picks up the $k_z$ component $\zeta = 2\pi c/d_3$, where $d_3$ is the spacing between Cu-O planes in the bilayer, and it occurs in the antibonding bilayer charge phasing channel.